\documentclass[letter]{jpsj2}

\title{Dephasing of coupled spin qubit system during gate operations 
due to background 
charge fluctuations}
\usepackage{graphicx}
\usepackage{dcolumn}
\usepackage{bm}

\author{
Toshifumi Itakura
\thanks{E-mail address: toshifumiitakura@scphys.kyoto-u.ac.jp}
}

\inst{Department of Physics, 
 Kyoto-University,
 Kyoto 606-8502, Japan \\
 TEL:075-753-3750
   }

\recdate{\today}

\abst{
It has been proposed that a quantum computer can 
be constructed based on electron spins in quantum dots 
  or based on a superconducting nanocircuit.
During two-qubit operations, the fluctuation of the coupling parameters
 is a critical factor.
One source of such fluctuation is the stirring of the background charges.
We focused on the influence of this fluctuation on a coupled spin qubit system.
The induced fluctuation in exchange coupling changes the amount 
of entanglement, fidelity, and purity. 
In our previous study, the background charge fluctuations were found
 to be an important channel of dephasing for a single Josephson qubit.
}

\kword{
quantum computation, dephasing, 
coupled spin qubits, background charge fluctuation
}

\begin{document}
\maketitle


Among the various proposals for quantum computation,
 quantum bits (qubits) in solid state materials,
           such as superconducting Josephson junctions
\cite{Nakamura}
           and quantum dots
\cite{Hayashi,Tanamoto,Loss},
           have the advantage of  scalability.
Proposals to implement a quantum computer using superconducting nanocircuits
    are proving to be very promising
    \cite{Makhlin,Makhlin_RMP,Averin,Mooij,Falci},
    and several experiments have already highlighted the quantum properties
    of these devices
    \cite{Bouchiat,Friedman,Aassime}.
Such a coherent-two-level system constitutes a qubit and
 the quantum computation can be carried out as the
        unitary operation functioning on the multiple qubit system.
Essentially, this quantum coherence must be maintained during computation.
However, dephasing is hard to 
     avoid due to the system's interaction with
     the environment.
The dephasing is characterized by the dephasing time $T_2$,
 and various environments can cause dephasing.

Background charge fluctuations (BCFs) 
         have been observed in various kinds of  systems
\cite{Devoret,Zorin,Martinis,Lyon}.
In nanoscale systems, BCFs are
     electrostatic potential fluctuations arising
     due to the dynamics of an electron, or hole,
     on a charge trap.     
In particular, the charges at charge traps 
     fluctuate
     with the  Lorentzian spectrum form,
     which is 	
     called
     random telegraph noise
     in the time domain
\cite{Lyon,Fujisawa_BC}.
The random 
      distribution of the positions of such dynamical charge traps
       and their time constants  
     leads to  BCFs or 1/{\it f} noise
\cite{BC}.
In solid-state charge qubits,
    these BCFs
    result in  a  dynamical electrostatic disturbance and
    hence, dephasing.
It should be noted that this dephasing process does not mean 
        the qubit 
        being entangled with the environment,
        but rather,
        that  
        the stochastically evolution of an external classical field
        is suppressing the density matrix elements 
        of the qubit after  averaging out over statistically distributed
         samples.

We had shown that BCFs are important channel of dephasing 
  for a single Josephson charge qubit system
\cite{Itakura_Tokura_PRB,Itakura_Tokura}.
In the present study, we investigate the effect of BCFs on the two-qubit-gate
    operation.
To construct a controllable quantum computer, one requires 
    the suppression of dephasing and
    accurate  universal quantum gate which  consists of
    single qubit operations and two-qubit operations.
Therefore, to address these manipulations, we examine 
     the dephasing of coupled qubit system,	
     which is experimentally current topic and is urgent to analyze what cause
     of dephasing is important in these systems.               


Recently, it has been shown that the interaction between electron spin
       in a quantum dot and environments is weak
\cite{Fujisawa_Nature,Fujisawa_Tokura},
 and one can expect a very long dephasing time of an electron spin.
Therefore, the proposal of quantum computer using electron spin
  in a quantum dot is promising
\cite{Loss}.
For the  electron spin qubit,
 however,
  the effects of ;
  the fluctuation of local magnetic field
\cite{Loss,DasSarma,Burkard_SO},
 and the spin-orbit interaction
\cite{Burkard_SO} are important.
Moreover, the fluctuation in the exchange coupling is important
 during the gate operation,
because, for two-qubit gate operation,
 one uses the exchange interaction between the two quantum dots.
This type of dephasing occurs when the charge of traps change the 
      coulomb energy of two-qubit state and
      fluctuate the exchange energy between the two qubits.
Therefore, we investigate the effect of BCFs
        on the two-qubit gate operation.

We examine the time evolution of two-qubit density matrix
  which obeys  the two-qubit Hamiltonian 
$  H = J (t) {\bf S_1} \cdot {\bf S_2},$
where $J(t)$ depends on time, and 
we neglect the magnetic field on  each qubit system.
We examine the amount of fluctuation in exchange coupling.(Fig1.)
\begin{figure}
\center
\includegraphics[width=0.8\linewidth]{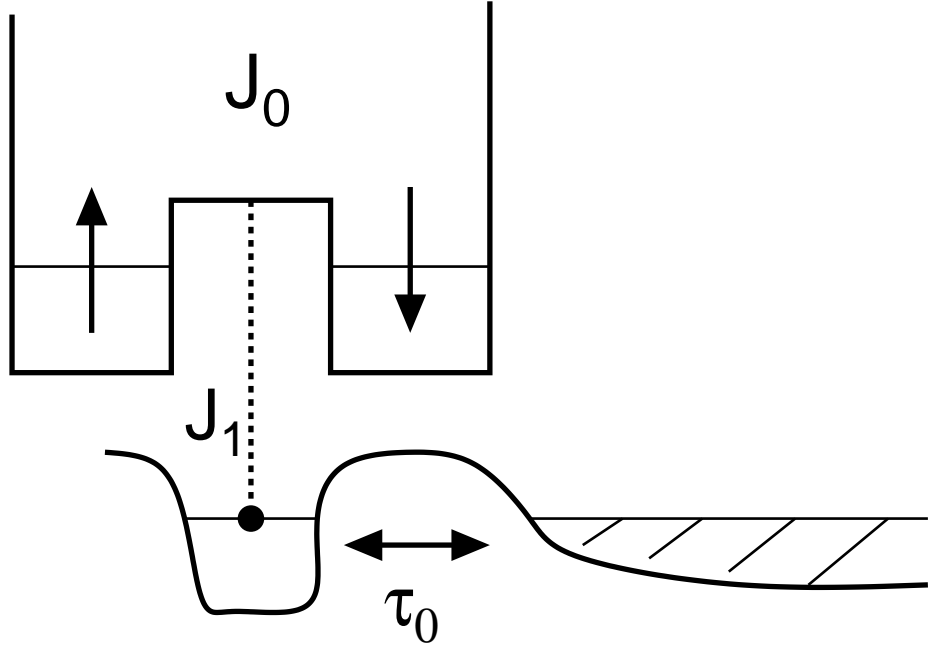}
\begin{minipage}[t]{8.5cm}
\caption{
Scheme of coupled qubit system and charge trap
}
\label{figure1}
\end{minipage}
\end{figure}	

We define the quantum variables of two qubits' coordinate as
     $\vec{r_1}$ and $\vec{r_2}$,
and define the environment variables of a charge trap and its near-by electron
     reservoir as $\vec{r_3}$ and $\vec{r_4}$.     
We define the Hamiltonian of two qubit system as double well
 in a two-dimensional layer ({\it z}=0) such that
$  H_0 = \sum_{i=1,2} h_i + C,
  h_i = \frac{1}{2m} p_i^2 + V(r_i),
  C = \frac{e^2}{|r_1-r_2|},
  V(x,y) = ( \frac{m \omega^2}{2}  \frac{1}{4a^2} (x^2-a^2)^2
            + \frac{m \omega^2}{2}   y^2 ), $
where $m$ is the effective mass of an electron in the quantum dot,
$\omega$ the confinement frequency of confinement potential,
and $2a$ the distance between two potential minima.
The exchange energy of electron spins between the
 dots is given by $ J_0=\frac{\hbar \omega}{\sinh (2 d^2 )} ( c (
       e^{-d^2} I_0 (  d^2) ) + \frac{3}{4} (1 + d^2)) $	
where $c=\sqrt{\pi/2}(e^2/a_B)/\hbar \omega$,
$I_0(x)$ is 0-th order Bessel function, and
$d=a/a_B$,  $a_B=\sqrt{\hbar/m \omega}$
\cite{Burkard}.
The interaction Hamiltonian is
$  V_1 = e^2 /\sqrt{|\vec{r_1} - \vec{r_3}|^2}
       - e^2 /\sqrt{|\vec{r_2} - \vec{r_4}|^2}
       \simeq e^2 /r  
       + 
      e^2 (( \vec{r_1} + \vec{r_2}) \cdot ( \vec{r_3} - \vec{r_4} ))/r^3, $
where $r \simeq \sqrt{|\vec{r_3}|^2} 
\simeq \sqrt{|\vec{r_4}|^2}$.
From the above calculation, the dynamic part of 
the interaction Hamiltonian between
   charge trap and qubit system is given by
$      V_1 (\vec{r_1},\vec{r_2}) 
      = 
      e^2 \frac{ ( x_1 + x_2 ) (x_3 - x_4) + ( y_1 + y_2) (y_3 - y_4)}{r^3}$,
where we set $z_1$ and $z_2$ to zero.
Using the Heitler-London approximation,
 we define, $|S \rangle$ and $|T \rangle$ to be singlet and triplet states
 such that,
$|S \rangle = ( |12 \rangle + |21 \rangle ) /\sqrt{2(1+S^2)},
|T \rangle =  ( |12 \rangle - |21 \rangle ) /\sqrt{2(1-S^2)},$
where, $S^2 = |\langle 1 | 2 \rangle|^2={\rm exp}(-2d^2)$ 
is an overlap integral and 
$|1 \rangle = \sqrt{\frac{m \omega}{\pi \hbar}} 
e^{-m  \omega ( (x-a)^2 +  y^2 )/ 2 \hbar}, 
 |2 \rangle = \sqrt{\frac{m \omega}{\pi \hbar}} 
 e^{-m \omega ( (x+a)^2 + y^2 )/2\hbar}.$
When the trap's dipole vector aligns in {\it y}-direction,
  the fluctuation of the exchange coupling does not exist.
Then, the amount of fluctuation due to a single charge trap is written by,
$ J_{1 {\rm single}}= \frac{\hbar \omega}{\sinh ( 2d^2  )} \frac{3}{2 d^2}
  ( \frac{e^2 a (x_3 - x_4)}{\hbar \omega r^3} )^2$
  \cite{Burkard}.
For many traps, the magnitude of fluctuation in the exchange coupling
    becomes as follows.
The dipole vector of BCFs is 
$p(\sin \theta_i \cos \phi_i ,\sin \theta_i \sin \phi_i, \cos \theta_i)$
at distance $r_i$,
where  $p$ is distance between a trap and 
 electron reservoir.
Then the magnitude of fluctuation is given by
$ J_1 = \sum J_{1 {\rm single}} 
 = \sum_i \frac{1}{\hbar \omega} \frac{3}{2} \frac{1}{d^2}
     (\frac{e^2 p \sin \theta_i \cos \phi_i a}{r{_i}^3})^2
     \frac{1}{\sinh (2 d^2)} 
     =   \frac{2 \pi}{3}
     \frac{e^4}{\hbar \omega}
     \frac{1}{\sinh (2 d^2 ) }
     \frac{N_i a^2 p^2}{ r_d^3 d^2}$,
 where  $N_i$ is the density of charge traps and 
$r_d$ is the distance where the dipole approximation becomes invalid
($a < r_d$).

Next, we examine the time evolution of qubit density matrix.
The time evolution of the qubit system is defined by the unitary operator
\begin{equation}
  U(t) = e^{ -\frac{i}{\hbar} \int_0^t J(\tau) 
      {\bf S_1} \cdot {\bf S_2} d \tau } .
\end{equation}
The exchange coupling operator is expressed by
 using permutation operator $P_{12}$ such that
\begin{eqnarray}
      {\bf S_1} \cdot {\bf S_2} = \frac{P_{12} + 1}{2}.
\end{eqnarray}
$P_{12}$ has following properties:
\begin{equation}
  \label{eqn:OPC}
  P^{2m}_{12} =1 , P^{2m+1}_{12}= P_{12},
\end{equation}
where $m$ is integer.
We assume the fluctuation in the BCF obeys the random telegraph-type noise.
$ J ( t ) = J_0 + J_1 X (t)$,
where $ X(t)$ takes the value of 1 or 0 with characteristic time $\tau$.
The density matrix at time {\it t} is given by
$     \rho (t) = \langle U (t) \rho
 (t=0) U^{ \dagger} (t) \rangle,  $
where $U(t)$ is unitary operator which describes the time evolution 
 in terms of the bases $| \downarrow \downarrow \rangle,
 | \downarrow \uparrow \rangle , | \uparrow \downarrow \rangle,
 | \uparrow \uparrow \rangle$ and
 $\langle \rangle$ is the ensemble average about the stochastic process
 \cite{Itakura_Tokura,Kubo,Papoulis}.
To accomplish two-qubit gate operation, the restricted subspace
 spanned by $|\uparrow \downarrow \rangle, | \downarrow \uparrow \rangle$
  is enough because other states are decoherence free from this type of dephasing.
In restricted sub-Hilbert space, the initial pure state is given by
$ | \Psi \rangle = \cos \frac{\theta}{2}|\uparrow \downarrow \rangle +
      \sin \frac{\theta}{2} e^{- i \phi} |\downarrow \uparrow \rangle$,
and the quantum state is represented
 by using a Bloch sphere as shown in Fig. 2.
\begin{figure}
\center
\includegraphics[width=1.0\linewidth]{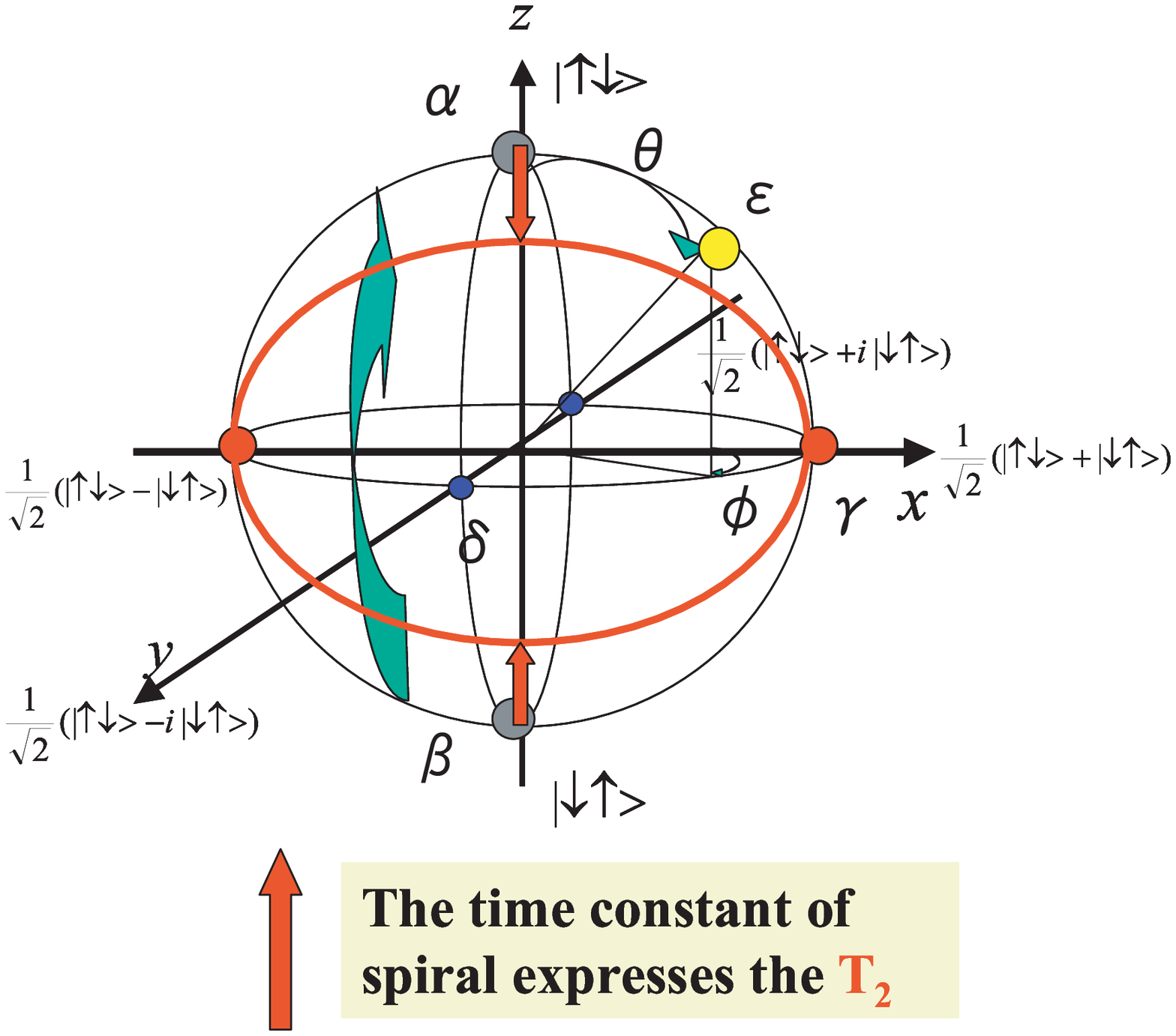}
\begin{minipage}[t]{8.5cm}
\caption{
Scheme of Bloch sphere with pointing moving quantum state.
$(\theta,\phi)$,$\alpha$=$(0,0)$,$\beta$=$(0,\pi)$,$\gamma$=$(\pi/2,0)$,$\delta$=($\pi/2,\pi$),$\epsilon$=($\pi/2,\pi/2$).
}
\label{figure2}
\end{minipage}
\end{figure}	
We obtain the density matrix at time {\it t} after taking the ensemble average
and calculate $T_2$ as had been done for a single qubit system
\cite{Itakura_Tokura,Kubo,Papoulis}.
The states on 
   one-dimensional line that connects two maximally entangled states 
   $(\frac{1}{2}(|\uparrow \downarrow \rangle + |\downarrow \uparrow \rangle)$ 
   and
   $(\frac{1}{2}(|\downarrow \uparrow \rangle - |\uparrow \downarrow \rangle)$.
   do not evolve during the gate operation.	
The distance from the origin of the Bloch sphere to quantum state 
  represents the purity.
The time evolution trace of the quantum state changes
 from sphere to ellipsoidal because of dephasing. 
The maximally entangled states
$|\Psi \rangle = (|\uparrow \downarrow \rangle
 \mp |\downarrow \uparrow \rangle)/\sqrt{2}$
  do not evolve in time,
since they are
  eigenstates of the Hamiltonian.
\begin{figure}
\center
\includegraphics[width=0.6\linewidth]{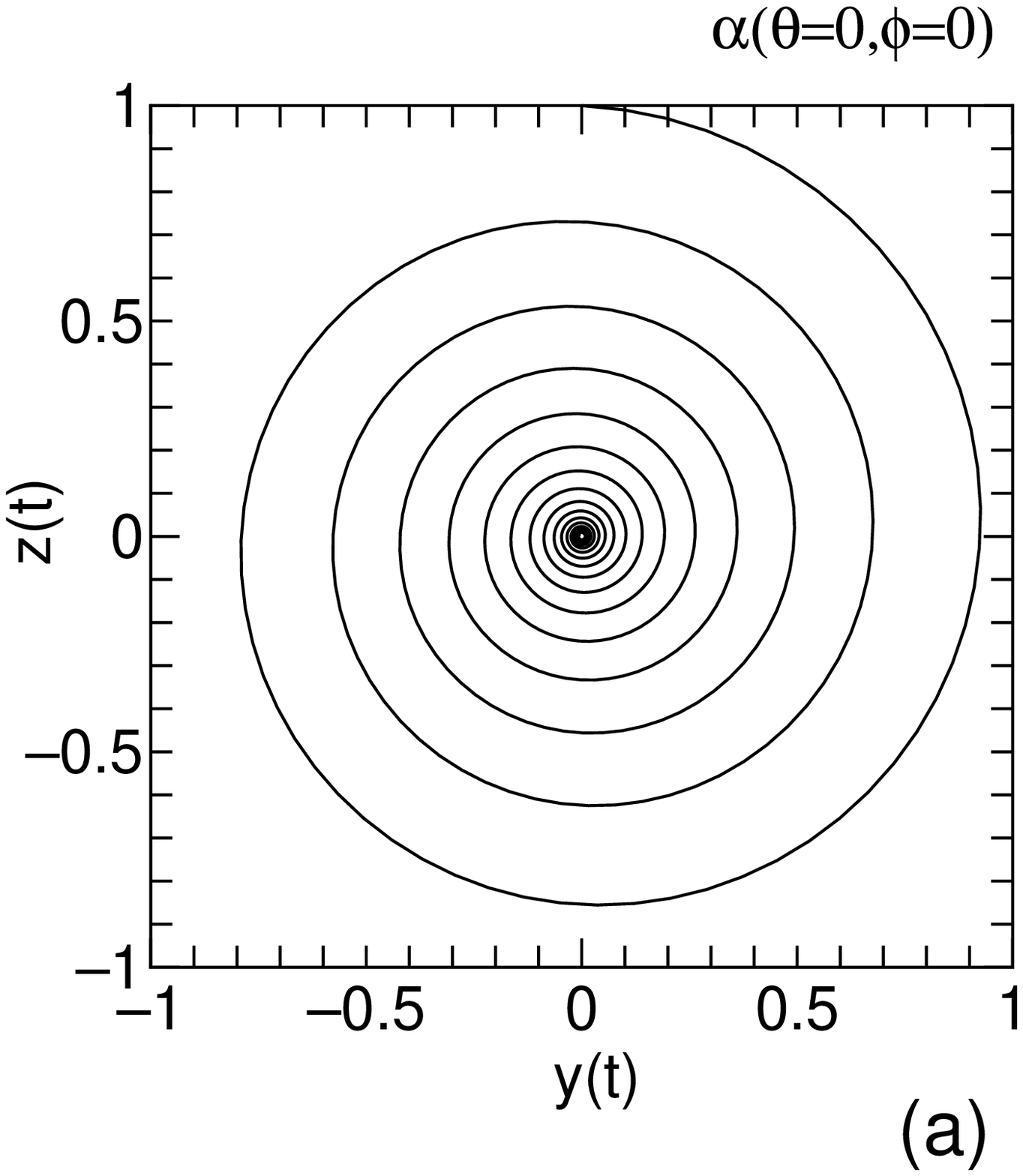}
\begin{minipage}[t]{8.5cm}
\end{minipage}
\center
\includegraphics[width=0.6\linewidth]{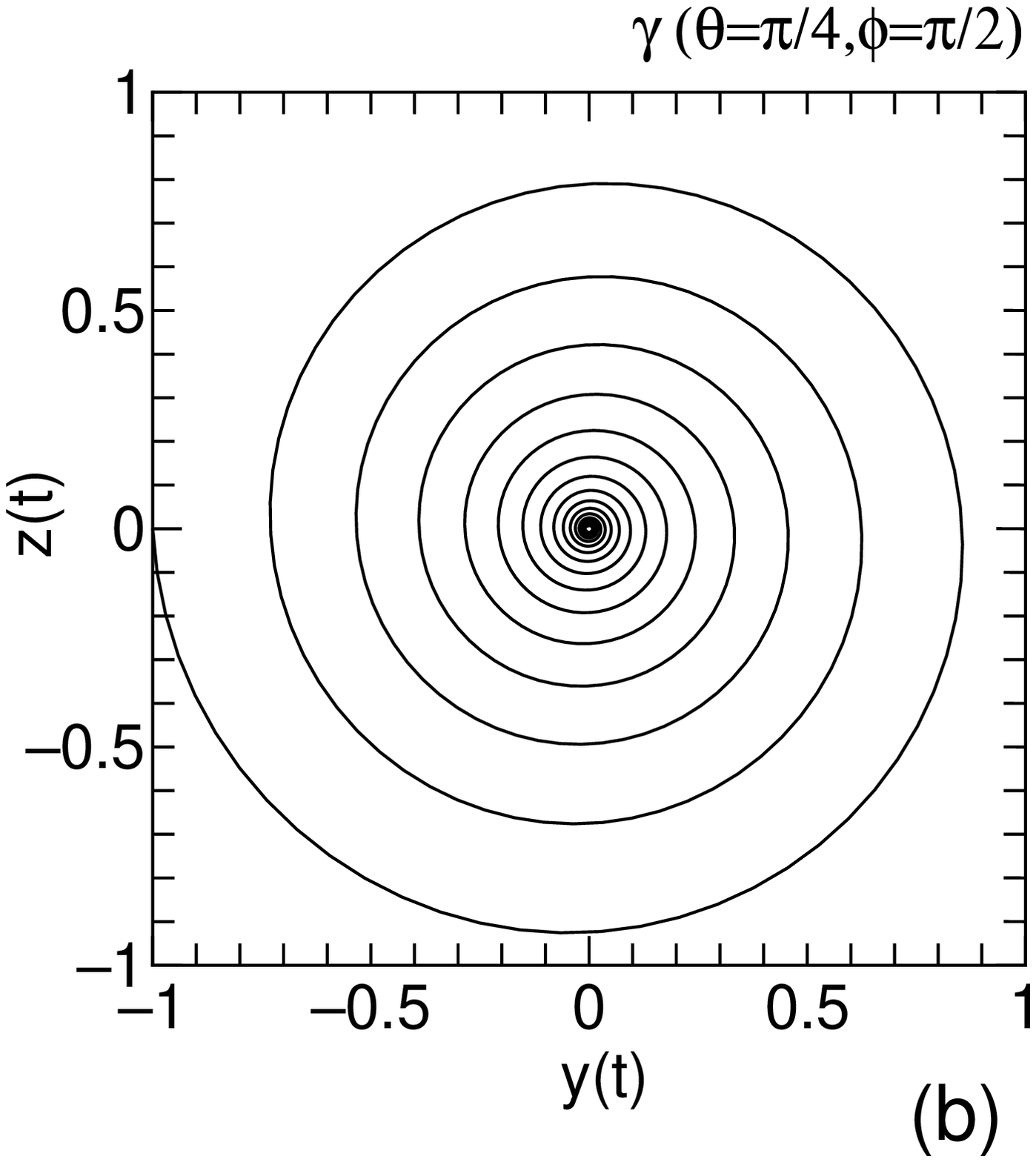}
\begin{minipage}[t]{8.5cm}
\caption{
Time depnedence of quantum state. 
(a) initial state $\alpha$=$(\theta=0,\phi=0)$.
(b) initial state $\beta$=$(\theta=\theta/4,\phi=\pi/2)$. 
}
\label{figure3}
\end{minipage}
\end{figure}
Fig. 3 (a),(b) shows time dependence of quantum state with some choice 
of initial condition.
We estimate $T_2^{-1}$ for many charge traps with different characteristic
   time as follows.
We assume the transition times between occupied and empty states are equal
 and the temperature dependence of transition time obeys 
 the thermal activation
 type 
 $\tau=Ae^{\frac{-W}{k_B T}}$,
 where $W$ is thermal activation energy and $A$ is characteristic time scale
 which is independent of temperature.
To estimate the effect of many charge traps,
  we average over the magnitude of fluctuation in the 
  exchange coupling and thermal activation energy.
Since the dephasing rate by a single trap is  $J_1^2 \tau \hbar$ for
  weak coupling and $\hbar /2 \tau$ for strong coupling
\cite{Itakura_Tokura_PRB,Itakura_Tokura},
 total $T_2^{-1}$ is given by     
$T_2^{-1} = \frac{3k_B T}{2W_0} J_1 $
where 
$1/W_0$ is distribution of thermal activation energy.
We define the gate operation time as $\tau_p =  \pi /J_0$.
Then, the gate quality factor is given by
$Q=T_2/\tau_p \simeq \frac{2W_0}{3 \pi k_B T} \frac{J_0}{J_1}$.

Next, we examine the quantum information quantities of
  the coupled qubit system.
First, we examine the criterion of entanglement.
This criterion comes from the negativity of minimum eigenvalue of
  a partially transposed qubit density matrix
\cite{Peres,Horodecki}.
\begin{figure}
\center
\includegraphics[width=0.8\linewidth]{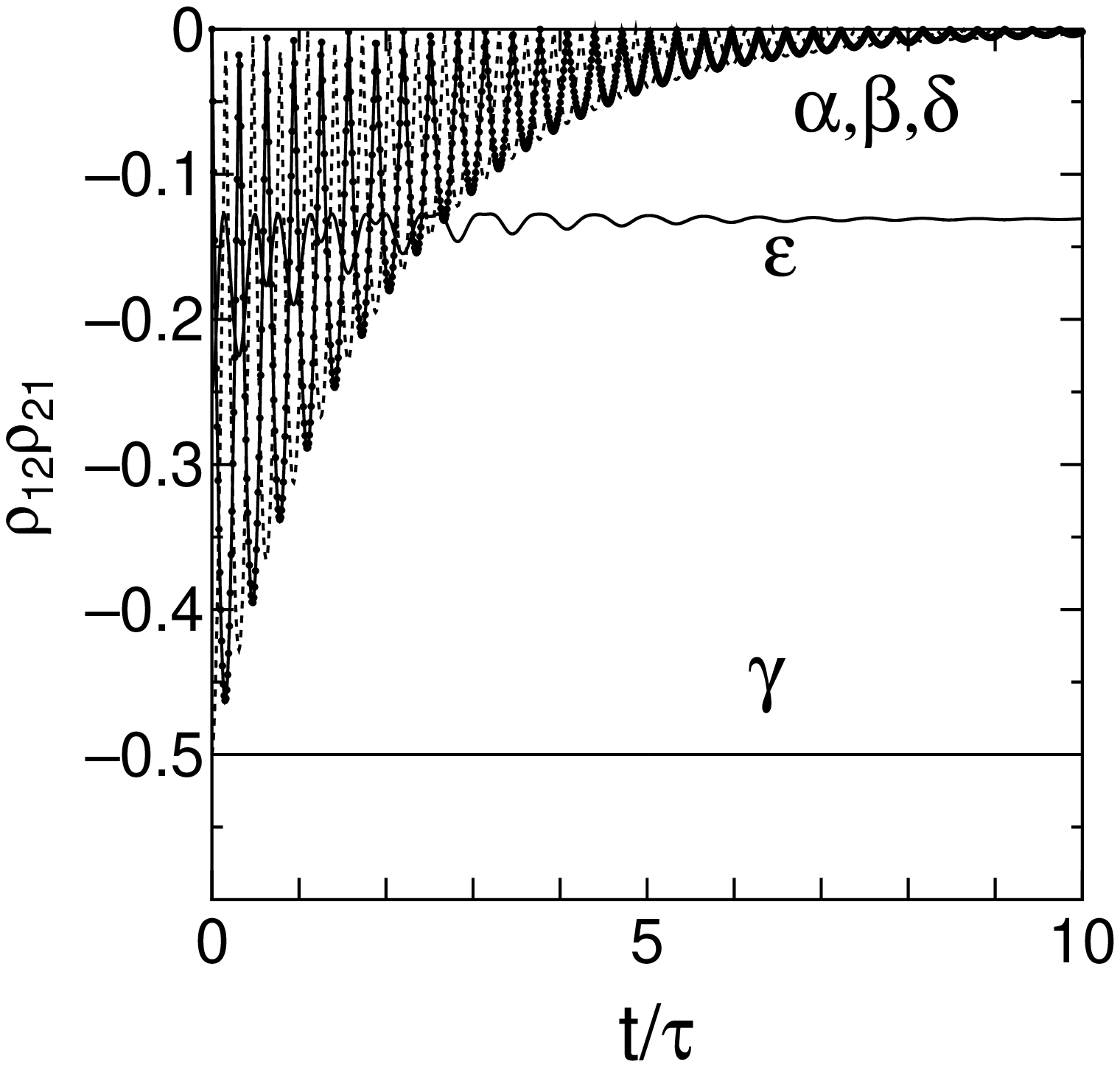}
\begin{minipage}[t]{8.5cm}
\caption{
Criterion of entanglement.}
\label{figure4}
\end{minipage}

\end{figure}
Fig. 4 shows time dependence of criterion of entanglement.
For Fig.4-6 we set $T_2/\tau=2$.
\begin{figure}
\center
\includegraphics[width=0.8\linewidth]{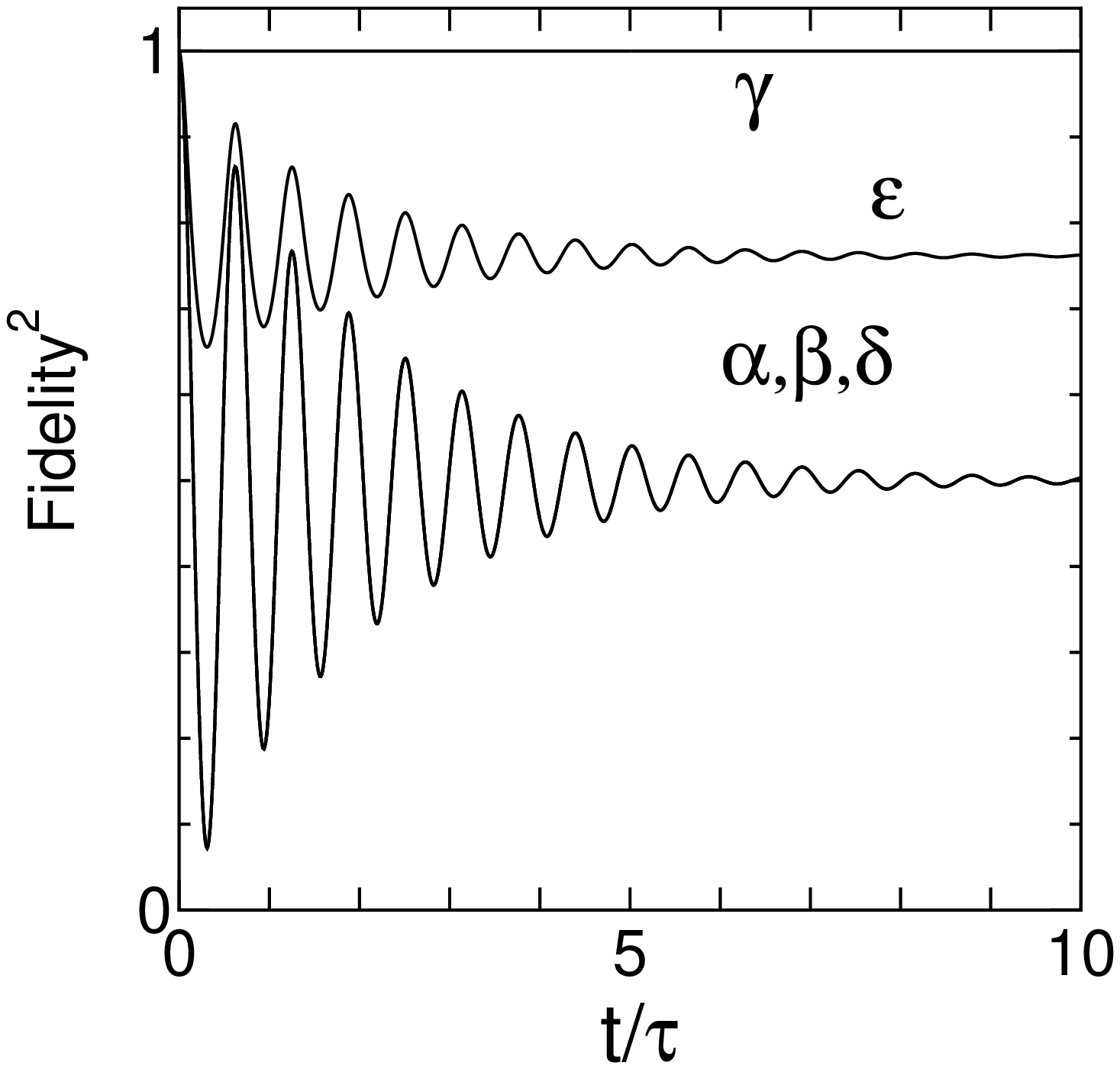}
\begin{minipage}[t]{8.5cm}
\caption{
Time dependence of fidelity.
}
\label{figure5}
\end{minipage}
\end{figure}
Fig. 5 shows time dependence of fidelity.
The minimum eigenvalue oscillates with time.
When $\sin \theta \cos \phi=0$, separable states, 	
$|\uparrow \downarrow \rangle, |\downarrow \uparrow \rangle$,
appear between entangled states during time evolution.
Whereas, the qubits are always entangled if 
$\sin \theta \cos \phi \ne 0$.
Next, we study the fidelity of $\rho(t)$ relative to $\rho(t=0)$,
defined by ${\cal F} (t) \equiv Tr [ \rho (t) \rho (0) ]$,
which shows accuracy of the quantum gate 
\cite{Poyatos,Nielsen}.
Starting from  the maximally entangled state, fidelity is 1.
For more general initial conditions, fidelity is given by
\begin{eqnarray}
 {\cal F}  &=& \frac{1}{2} + \frac{1}{2}
  \cos ( \frac{1}{\hbar} J_0 t)  e^{- t/T_2 } \cos^2 \theta \nonumber \\
         &+& 
            \frac{1}{2} \cos ( \frac{1}{\hbar} J_0 t)  e^{- t/T_2 }
             \sin^2  \theta  \sin^2 \phi \nonumber \\
  &+& \frac{1}{2} \sin^2 \theta \cos^2 \phi. 
\end{eqnarray}
Finally, we examine the purity (${\cal P} = Tr ( \rho (t)^2 )$),
 which is related to linear entropy
of qubit system as $S_{lin} = 1 - {\cal P}$
\cite{Poyatos,Shimizu}.
The analytical expression of purity is 
\begin{eqnarray}
 {\cal P} &=& \frac{1}{2}+\frac{1}{2} \sin^2  \theta \cos^2 \phi \nonumber \\
       &+& \frac{1}{2} 
        ( \sin^2  \theta \sin^2 \phi + \cos^2  \theta ) e^{-2t/T_2}.
\end{eqnarray}

\begin{figure}
\center
\includegraphics[width=0.8\linewidth]{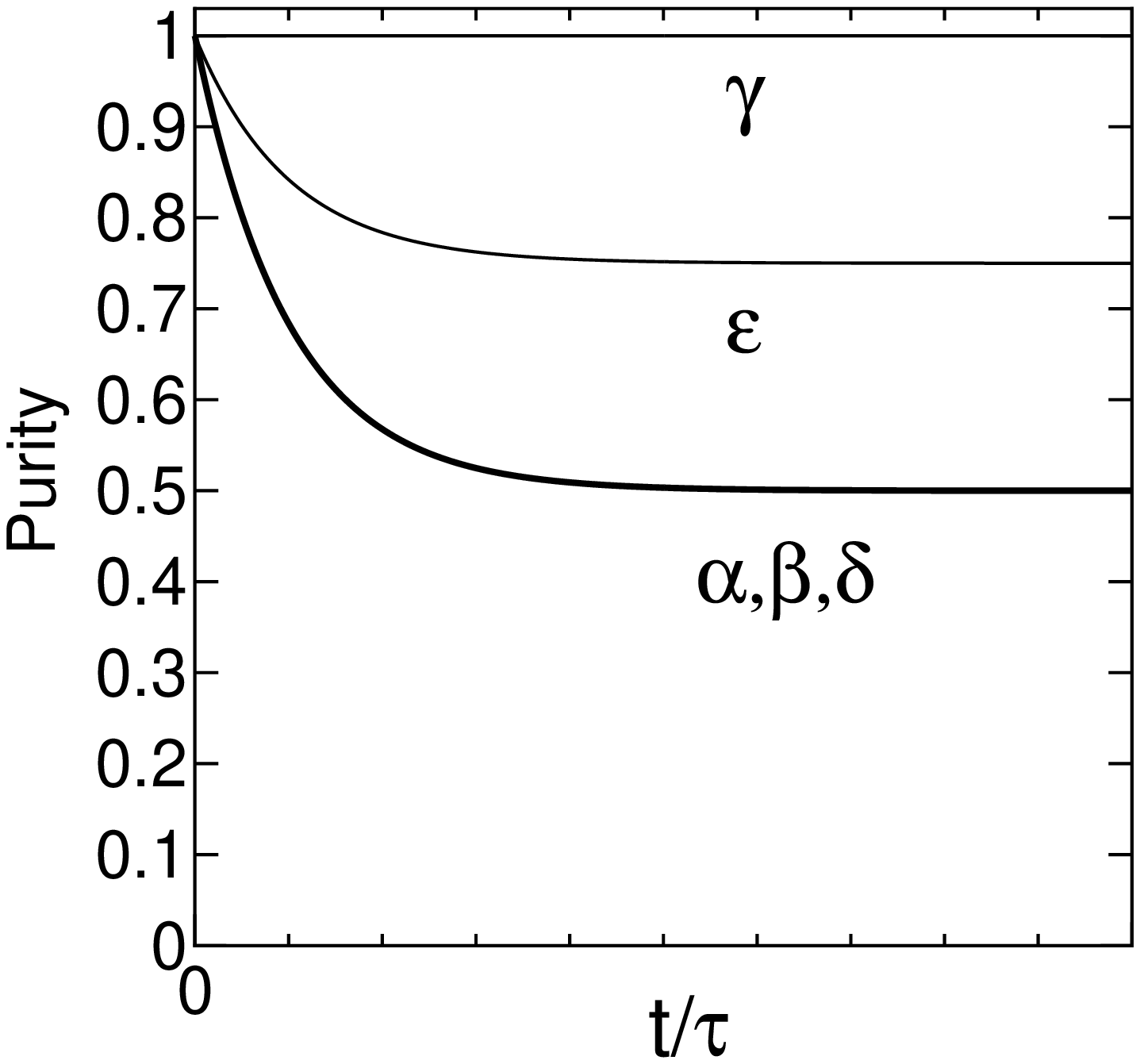}
\begin{minipage}[t]{8.5cm}
\caption{
Time dependence of purity.
}
\label{figure6}
\end{minipage}
\end{figure}	
Figure 6 shows the time dependence of 
purity.
The purity decreases monotonically,
which means  that the entropy of the qubit increases due to the dephasing.
The initially pure qubit state becomes a mixed state,
namely, ${\cal P} (t=0)=1$ becomes 
${\cal P} (t = \infty) = \frac{1}{2} + \frac{1}{2} \sin^2 \theta \cos^2 \phi$.

%

{\bf Acknowledgements} 
The authors thank Toshiaki Hayashi, 
Hayato Nakano, Toshimasa Fujisawa, Gerrit E. W. Bauer
 and Seigo Tarucha
for their advice and stimulating discussions.
This work is supported by the Grant-in-Aid for the 21st Century COE "Center for Diversity and Universality in Physics" from the Ministry of Education, Culture, Sports, Science and Technology (MEXT) of Japan.

\end{document}